\documentclass[12pt]{iopart}
\usepackage{graphicx}
\usepackage{cite}
\usepackage{amsfonts}
\usepackage{array}
\usepackage{booktabs}

\begin{document}
\title[Epidemic spreading with long-range infections and incubation times]
      {Epidemic spreading with long-range infections and incubation times}
\author{Julian Adamek, Michael Keller, Arne Senftleben, and \\Haye Hinrichsen}

\address{Fakult\"at f\"ur Physik und Astronomie, 
         Universit\"at W\"urzburg,\\ D-97074 W\"urzburg, Germany}

\begin{abstract}
The non-equilibrium phase transition in models for epidemic spreading with long-range infections in combination with incubation times is investigated by field-theoretical and numerical methods. In this class of models the infection is assumed to spread isotropically over long distances $r$ whose probability distribution decays algebraically as $P(r) \sim r^{-d-\sigma}$, where $d$ is the spatial dimension. Moreover, a freshly infected individual can infect other individuals only after a certain incubation time $\Delta t$, which is distributed probabilistically as $P(\Delta t) \sim (\Delta t)^{-1-\kappa}$. Tuning the balance between spreading and spontaneous recovery one observes a continuous phase transition from a fluctuating active phase into an absorbing phase, where the infection becomes extinct. Depending on the parameters $\sigma$ and $\kappa$ this transition between spreading and extinction is characterized by continuously varying critical exponents, extending from a mean-field regime to a phase described by the universality class of directed percolation. Specifying the phase diagram in terms of $\sigma$ and $\kappa$ we compute the critical exponents in the vicinity of the upper critical dimension $d_c=\sigma(3-\kappa^{-1})$ by a field-theoretic renormalization group calculation and verify the results in one spatial dimension by extensive numerical simulations.
\end{abstract}

\submitto{Journal of Statistical Mechanics: Theory and Experiment}
\maketitle
\def\xvec{{\mathbf{x}}}
\def\kvec{{\mathbf{k}}}
\def\text#1{\mbox{#1}}
\parskip 1mm 
%
\section{Introduction}
\label{intro}
%
Epidemic spreading within a population of spatially distributed individuals is a complex phenomenon which depends on many parameters such as the density and the susceptibility of individuals, the specific transport mechanism, incubation times and memory effects like immunisation. Depending on the interplay of these factors an infectious disease may either spread over the population or disappear after some time. In order to understand the transition between survival and extinction of the pathogen various models with different degrees of simplification have been discussed in the literature~\cite{Bailey75,Mollison77,AndersonMay91,Daley99,Anderson00}. The interest of physicists and mathematicians in models of epidemic spreading is mainly motivated by the observation that in many cases the transition is continuous  and characterized by {\em universal} properties determined by the symmetries of the model which are independent of microscopic details. This allows one to classify phase transitions in epidemic models into universality classes characterized by certain critical exponents and universal scaling functions~\cite{MarroDickman,OdorReview,Lubeck}. Typically these classes correspond to different underlying field theories that describe the long-range properties of the dynamics at the transition.

Directed percolation (DP), also known as contact process, is probably the simplest and best-studied universality class of transitions in epidemic models~\cite{Kinzel85,Hinrichsen00}. This class is characterized by three independent critical exponents $\beta,\nu_\perp$, and $\nu_\parallel$. In the stationary active phase these exponents are related to the density of infected individuals $\rho \sim (p-p_c)^\beta$, the spatial correlation length $\xi_\perp\sim(p-p_c)^{-\nu_\perp}$, and the temporal correlation length $\xi_\parallel\sim(p-p_c)^{-\nu_\parallel}$, where $p$ is the infection probability and $p_c$ its critical threshold. Directed percolation corresponds to a specific type of field theory which made its first appearance in the context of high-energy physics~\cite{Reggeon1,Reggeon2,Reggeon3}. This observation led Janssen and Grassberger to their celebrated conjecture that any model which complies with the symmetry properties of this field theory should belong to the DP class~\cite{Janssen81,Grassberger82}. One of these conditions is that the infections have to be \textit{local} in space and time.

In order to study more realistic models which are still simple enough to be treated analytically but sufficiently complex to exhibit non-DP behavior, two important generalisations have been investigated. The first one is an epidemic process with long-range spreading, where the infectious disease is transported in random directions over long distances~\cite{Mollison77}. Motivated by empirical data the spreading distances $r$ in these generalized models are assumed to be distributed as a power law
\begin{equation}
P(r) \sim r^{-d-\sigma}\,, \qquad \qquad (\sigma>0)
\end{equation}
where $\sigma$ is a control exponent. In the literature such algebraically distributed long-range displacements are known as \textit{L{\'e}vy flights}~\cite{Levy} and have been studied extensively e.g. in the context of anomalous diffusion~\cite{Fogedby94}. In the present context of epidemic spreading long-range flights do not destroy the transition, instead they change the critical behavior provided that $\sigma$ is sufficiently small. More specifically, it was observed both numerically and in mean field approximations that the critical exponents change \textit{continuously} with~$\sigma$~\cite{Grassberger86,MarquesFerreira94,Albano96}. As a major breakthrough, Janssen {\it et al} introduced a renormalizable field theory for epidemic spreading transitions with spatial L{\'e}vy flights~\cite{JOWH99}, computing the critical exponents to one-loop order. Because of an additional scaling relation only two of the three exponents were found to be independent. These results were confirmed numerically by Monte Carlo simulations~\cite{HinrichsenHoward99}. As a further step towards more realistic models, Brockmann and Geisel~\cite{TheoGeisel} recently studied epidemic spreading by L{\'e}vy flights on a \textit{inhomogeneous} support in the supercritical phase.

The second generalization introduces a similar long-range mechanism in \textit{temporal} direction. Such `temporal' L{\'e}vy flights can be interpreted as incubation times $\Delta t$ before an infected individual can subsequently infect others. As in the first case, these incubation times are assumed to be algebraically distributed as
\begin{equation}
P(\Delta t) \sim \Delta t^{-1-\kappa}\,, \qquad\qquad  (\kappa>0)
\end{equation}
where $\kappa$ is a control exponent. However, unlike spatial L{\'e}vy flights, which take place equally distributed in all directions, such temporal L{\'e}vy flights have to be directed forward in time. A field-theoretic renormalization group calculation by Jim{\'e}nez-Dalmaroni and Cardy~\cite{DalmaroniCardy} allows the critical exponents to be computed to one-loop order. Again only two of the three exponents turn out to be independent because of an additional scaling relation.

In the present paper we investigate the mixed case of epidemic spreading by spatial L{\'e}vy flights {\it combined} with algebraically distributed incubation times. As we will show, the corresponding field theory renders two additional scaling relations, hence only one of the three exponents is independent. We present a phase diagram in terms of the control exponents $\sigma$ and $\kappa$, specifying the mean-field phase, a DP phase, two phases corresponding to the previously studied cases, and a novel fluctuation-dominated phase describing the mixed case in which the critical exponents are computed by a field-theoretic renormalization group calculation to one-loop order. Investigating several cross sections of the phase diagram by high-precision Monte Carlo simulations we demonstrate that the field-theoretic predictions are in fair agreement with numerical results.

\section{Modelling long-range spreading and incubation times}
\label{model}
%

Ordinary DP with short-range interactions is described by the Langevin equation~\cite{Janssen81}
\begin{equation}
\label{DPLangevin}
\tau \partial_t\rho = a \rho - b \rho^2 + D \nabla^2 \rho + \xi(\xvec,t) \,,
\end{equation}
where $\rho(\xvec,t)$ is a coarse-grained density of infected individuals and $\tau,a,b,D$ are certain coefficients ($\tau$ is introduced for later convenience and may be set to 1). The Gaussian noise $\xi(\xvec,t)$ describes density fluctuations. Its intensity varies with the density of active sites, as described by the correlation function
\begin{equation}
\label{Noise}
\langle \xi(\xvec,t) \xi(\xvec',t')\rangle = 2c \, \rho(\xvec,t) \, \delta^d(\xvec-\xvec')\delta(t-t')\,.
\end{equation}
Following Refs.~\cite{JOWH99,HinrichsenHoward99,DalmaroniCardy}, the generalized case of algebraically distributed long-range infections and incubation times is described by the Langevin equation
\begin{equation}
\label{FullLangevin}
(\tau \partial_t + \tilde{\tau} \tilde{\partial}_t^\kappa) \rho = 
a \rho - b \rho^2 + (D \nabla^2 + \tilde{D}\tilde{\nabla}^\sigma) \rho + \xi(\xvec,t)
\end{equation}
with the same noise as in Eq.~(\ref{DPLangevin}). Here $\tilde{\partial}_t^\kappa$ and $\tilde{\nabla}^\sigma$ are linear non-local operators acting on $\rho(\xvec,t)$ by
\begin{equation}
\label{IntegralTime}
\tilde{\partial}_t^\kappa \, \rho(\xvec,t) = \frac{1}{\mathcal{N}_\parallel}
\int_0^{\infty} {\rm d}t' \, {t'}^{-1-\kappa} [\rho(\xvec,t)-\rho(\xvec,t-t')]
\qquad (0 < \kappa < 1)
\end{equation}
and
\begin{equation}
\label{IntegralSpace}
\tilde{\nabla}^\sigma \,\rho(\xvec,t) = \frac{1}{\mathcal{N}_\perp}
\int {\rm d}^dx' \, |\xvec'|^{-d-\sigma} [\rho(\xvec+\xvec',t)-\rho(\xvec,t)]
\qquad (0 < \sigma < 2)
\end{equation}
where $\mathcal{N}_\parallel=-\Gamma(-\kappa)$ and $\mathcal{N}_\perp=-2\cos(\pi\sigma/2)\Gamma(-\sigma)$ are appropriate normalization constants. Acting on a plane wave these operators fulfill the properties
\begin{eqnarray}
\tilde{\partial}_t^\kappa \, e^{i\omega t} &=& (i\omega)^\kappa \,  e^{i\omega t} \,, \\
\tilde{\nabla}^\sigma \, e^{i \kvec \cdot \xvec} &=& -|\kvec|^\sigma \, e^{i \kvec \cdot \xvec} 
\end{eqnarray}  
so that they can be interpreted as fractional derivatives~\cite{Fogedby94}. As an important difference the temporal operator is \textit{directed} in time and thus brings down the nonsymmetric factor $(-i\omega)^\kappa$ in front of the exponential, while the \textit{undirected} spatial operator renders the factor $-|\kvec|^\sigma$ which is symmetric under reflections. We note that the operator $\tilde{\nabla}^2$ is still non-local and hence differs from the ordinary Laplacian $\nabla^2$ although both operators act in the same way on a plane wave. Similarly, $\tilde{\partial}_t^1$ is non-local and differs from the ordinary local derivative $\partial_t$.

Since the integral kernels in Eqs.~(\ref{IntegralSpace}) and~(\ref{IntegralTime}) would diverge for $t'\to 0$ and $\xvec'\to 0$ one has to introduce lower cutoffs for the spreading distances in space and time. These cutoffs induce short-range components which are accounted for by retaining the local operators $\partial_t$ and $\nabla^2$ in Eq.~(\ref{FullLangevin}). Even if these terms were not initially included, they would be generated under renormalization of the theory.

A dimensional analysis of Eq.~(\ref{FullLangevin}) immediately yields the mean field exponents
\begin{equation}
\label{MFExponents}
\beta^{\rm MF}=1\,, \quad
\nu_\perp^{\rm MF}=\sigma^{-1}\,, \quad
\nu_\parallel^{\rm MF}=\kappa^{-1}\,, \quad
\end{equation}
and the upper critical dimension
\begin{equation}
d_c = 3\sigma - \frac{\sigma}{\kappa}
\end{equation}
below which fluctuation effects are expected to become relevant. These exponents vary continuously with $\sigma$ and $\kappa$. As expected, for $\sigma\to 2$ and $\kappa\to 1$ they cross over smoothly to the mean field exponents of DP in $d \geq 4$ spatial dimensions.

\section{Field-theoretic renormalization group calculation}
\label{fieldtheory}
%
\subsection{Field-theoretic action}
Starting point of a field-theoretic renormalization group calculation is an effective action~$S$ which is defined as the partition sum over all realizations of the density field $\psi(\xvec,t):=\rho(\xvec,t)$ and the noise field $\xi(\xvec,t)$, constrained to solutions of the Langevin equation~(\ref{FullLangevin}). Integrating out the noise by introducing a Martin-Siggia-Rosen response field $\bar{\psi}(\xvec,t)$ and symmetrizing the coefficients of the cubic terms by rescaling the field $\psi$ and $\bar{\psi}$ one arrives at the field theoretic action
\begin{equation}
S[\psi,\bar{\psi}]=S_0[\psi,\bar{\psi}]+S_{\rm int}[\psi,\bar{\psi}]
\end{equation} 
consisting of a free contribution
\begin{equation}
S_0[\psi,\bar{\psi}]\,=\,\int\,{\rm d}^dx\,{\rm d}t \,\, \bar{\psi}
(\tau \partial_t + \tilde{\tau} \tilde{\partial}_t^\kappa
-a-D\nabla^2 -\tilde{D} \tilde{\nabla}^\sigma)
\psi
\end{equation} 
and an interaction part
\begin{equation}
S_{\rm int}[\psi,\bar{\psi}]\,=\,g\,\int\,{\rm d}^dx\,{\rm d}t \,\, 
(\bar{\psi}\psi^2-\bar{\psi}^2\psi)
\end{equation} 
with the coupling constant $g=\sqrt{bc}$. 

\begin{figure}
\centerline{\includegraphics[width=106mm]{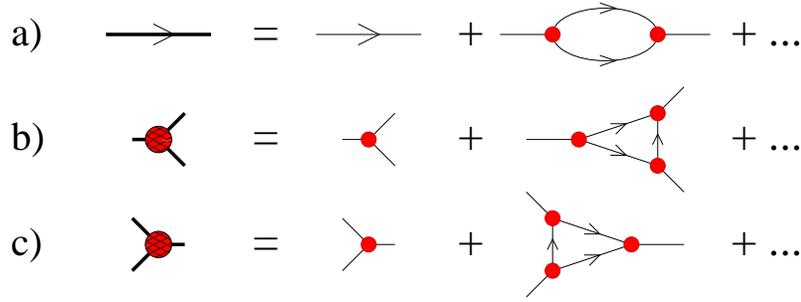}}
\caption{\label{FIGFEYNMAN} \small
Feynman diagrams of directed percolation. The figure shows the one-loop expansion of a) the propagator and b)-c) the cubic vertices. The thin lines represent the free propagator $G_0(\kvec,\omega)$ directed in time according to the arrow while the small circles denote bare cubic vertices $\pm g$. 
}
\end{figure}

It is important note that the additional long-range operators $\tilde{\partial}_t^\kappa$ and $\tilde{\nabla}^\sigma$ appear only in the \textit{free} part of the action. Therefore, the structure of the Feynman diagrams in a loop expansion is exactly the same as in DP (see Fig.~\ref{FIGFEYNMAN}), the only difference being that the free propagator in momentum space $G_0^{\rm DP}(\kvec,\omega) =(D k^2-a-i\tau\omega)^{-1}$ is replaced by the generalized counterpart
\begin{equation}
G_0(\kvec,\omega) = \Bigl[ D k^2 + \tilde{D}k^\sigma -a-i\tau \omega-\tilde{\tau}(i\omega)^\kappa \Bigr]^{-1} \,.
\end{equation} 
%
%
%

\noindent
Apart from this modification the loop integrals therefore have exactly the same structure as in DP. In particular, the so-called rapidity reversal symmetry
\begin{equation}
\psi(\kvec,\omega)\to-\bar{\psi}(\kvec,-\omega)\,\qquad
\bar{\psi}(\kvec,\omega)\to-\psi(\kvec,-\omega)
\end{equation} 
is still valid. As a consequence the fields $\psi$ and $\bar{\psi}$ scale identically and the cubic vertices shown in Fig.~\ref{FIGFEYNMAN}b-\ref{FIGFEYNMAN}c renormalize exactly in the same way.

Wilson's momentum shell renormalization procedure consists of two steps. At first the action is rescaled by
\begin{equation}
\xvec \to b \xvec,\quad
t \to b^z t,\quad
\psi \to b^\chi \psi ,\quad
\bar{\psi} \to b^\chi \bar{\psi}
\end{equation} 
where $b=1-\ell$ with $0 < \ell \ll 1$ is a dilatation factor while $z=\nu_\parallel/\nu_\perp$ and $\chi=-\beta/\nu_\perp$ are critical exponents. Introducing an upper cutoff $\Lambda$ in momentum space this scaling transformation inflates the cutoff by $\Lambda\to\Lambda/b$. In the second step of Wilson's renormalization procedure, the newly arising short wavelengths within the momentum shell $\Lambda < |\kvec| < \Lambda/b$ are integrated out in the loop expansion and expanded to lowest order in $\kvec$ and $\omega$, renormalizing the coefficients of the propagator and the vertices. In the present case, shell integration to one-loop order renormalizes the propagator and the coupling constant by
\begin{equation}
G^{-1}(\kvec,\omega) \to G^{-1}(\kvec,\omega) + \ell\,L(\kvec,\omega)\,,\qquad
g \to g-\ell\, V
\end{equation} 
where
\small
\begin{eqnarray}
\label{LIntegral}
L(\kvec,\omega) &=& 2g^2 \int_{|\kvec|=\Lambda}
\frac{{\rm d}^dk'\,{\rm d}\omega'}{(2\pi)^{d+1}}
G_0\Bigl(\frac{\kvec}{2}+\kvec',\frac{\omega}{2}+\omega'\Bigr)
G_0\Bigl(\frac{\kvec}{2}-\kvec',\frac{\omega}{2}-\omega'\Bigr)
\\
\label{VIntegral}
V&=&8g^3\int_{|\kvec|=\Lambda}
\frac{{\rm d}^dk'\,{\rm d}\omega'}{(2\pi)^{d+1}}\,
G_0^2(\kvec',\omega')G_0(-\kvec',-\omega')\,.
\end{eqnarray}
\normalsize
are the shell integrals corresponding to the one-loop diagrams shown in Fig.\ref{FIGFEYNMAN}. At the fixed point of rescaling combined with shell integration the system is assumed to be critical and the two exponents $\chi$ and $z$ can be calculated. 

\subsection{Directed percolation}
%
Before calculating the critical exponents for generalized epidemic spreading with spatio-temporal long-range interactions let us briefly recall Wilson's renormalization group approach applied to DP~\cite{Wijland98}. Setting $\tilde{\tau}=\tilde{D}=0$ the action involves four coefficients $g,a,\tau$, and $D$. Under combined action of rescaling and shell integration these coefficients change as
\begin{eqnarray}
\label{FlowEquations}
\partial \ell \ln g    &=& 3\chi + d + z - L_g\\
\partial \ell \ln a    &=& 2\chi + d + z - L_a\\
\partial \ell \ln \tau &=& 2\chi + d - L_\tau\\
\partial \ell \ln D    &=& 2\chi + d + z - 2 - L_D
\end{eqnarray} 
with the loop corrections
\footnotesize
\begin{equation}
L_g=\frac{V}{g},\quad 
L_a=\frac{L(0,0)}{a},\quad
L_\tau=\left.-\frac{i}{\tau}\,\frac{\partial L(0,\omega)}{\partial\omega}\right|_{\omega=0},\quad
L_D=\left.-\frac{1}{2D}\,\frac{\partial^2L(k,0)}{{\partial k}^2}\right|_{k=0}
\end{equation} 
\normalsize
In the case of DP the shell integrals can be computed, leading to
\begin{equation}
L_g =\frac{2g^2\Lambda^d K_d}{\tau (D\Lambda^2-a)^2},\quad
L_a = \frac{g^2 \Lambda^d K_d}{a \tau (D\Lambda^2-a)},\quad
L_\tau=\frac{L_g}{4},\quad
L_D=\frac{L_g}{8}. 
\end{equation} 
where $K_d=\frac{1}{2^{d-1}\pi^{d/2}\Gamma(d/2)}$ is the surface of a $d$-dimensional unit sphere divided by $(2\pi)^d$.
Inserting these results the flow equations have the fixed point
\begin{equation}
L_a^*=2+\frac{\epsilon}{12},\qquad
L_g^*=\frac{2\epsilon}{3},\qquad
\chi=-2+\frac{7\epsilon}{12},\qquad
z=2-\frac{\epsilon}{12}\,,
\end{equation} 
where $\epsilon=d_c-d$ denotes the distance from the upper critical dimension $d_c=4$. Assuming $D$ and $\tau$ to be stationary and expanding these relations to first order in $\epsilon$ this fixed point corresponds to
\begin{equation}
(g^*)^2=\frac{D^2 \tau}{3 K_d}\epsilon\,,\qquad 
a^* = \frac{D \Lambda^2}{6} \epsilon\,.
\end{equation} 
In order to compute the third critical exponent one has to consider the flow equations for $a$ and $g$ in the vicinity of the fixed point. To this end one has to linearize the flow equations and to compute the eigenvalues of the matrix
\begin{equation}
\label{MMatrix}
M=\left(
\begin{array}{ll}
\partial_g g (3\chi+d+z-L_g) &
\partial_g a (2\chi+d+z-L_a) \\
\partial_a g (3\chi+d+z-L_g) &
\partial_a a (2\chi+d+z-L_a)
\end{array}
\right)_{g=g^*,\, a=a^*}
\end{equation} 
evaluated at the fixed point and to select the positve eigenvalue $2-\epsilon/4$, which is equal to $\nu_\perp^{-1}$. To first order in $\epsilon$ we therefore arrive at the well-known result for DP
\begin{equation}
\beta = 1 - \frac{\epsilon}{6}+{\rm O}(\epsilon^2),\quad
\nu_\parallel = 1 + \frac{\epsilon}{12}+{\rm O}(\epsilon^2) ,\quad
\nu_\perp = \frac12+\frac{\epsilon}{16}+{\rm O}(\epsilon^2) .
\end{equation} 
%
%
\subsection{Spatio-temporal long range interactions}
%
%
After summarizing the known case of DP let us now discuss the generalized model with spatio-temporal long-range interactions. As discussed above, by introducing long-range flights and incubation times we modify solely the free propagator while the structure of the loop integrals (\ref{LIntegral}) and (\ref{VIntegral}) remains unchanged. However, expanding these integrals as a Taylor series around $k=\omega=0$ one always obtains integral powers of $k$ and $\omega$ while fractional terms of the form $k^\sigma$ or $(i\omega)^\kappa$ are \textit{not} generated. This means that the fractional operators $\tilde{\partial}_t$ and $\tilde{\nabla}^\sigma$ do not renormalize themselvses, instead they renormalize their short-range counterparts $\partial_t$ and $\nabla^2$. Assuming this observation to hold to any order of perturbation theory, the RG flow equations for the coefficients of the long-range operators are determined exclusively by the rescaling-part
\begin{eqnarray}
\label{eqtaus}
\partial_\ell \ln \tilde{\tau} &=& 2\chi+d+z-\kappa z\,,\\
\label{eqDs}
\partial_\ell \ln \tilde{D} &=& 2\chi+d+z-\sigma \,.
\end{eqnarray}
At the fixed point the right-hand sides of these equations are zero, implying two exact scaling relations, namely, the temporal relation $2\chi = \sigma-d-z$ and the spatial relation $2 \chi=(\kappa-1)z-d$. In terms of the standard exponents $\beta,\nu_\perp,\nu_\parallel$ these scaling relations read
\begin{eqnarray}
\label{TScaling}
2\beta + (\sigma-d)\nu_\perp-\nu_\parallel&=&0\,,\\
\label{SScaling}
2\beta - d\nu_\perp + (\kappa-1)\nu_\parallel&=&0\,.
\end{eqnarray} 
Hence two of the three critical exponents are already determined by trivial renormalization of the long-range operators. The first one is the exponent for the temporal decay of the density
\begin{equation}
\delta = \frac{\beta}{\nu_\parallel} = \frac{d\kappa + \sigma - \kappa \sigma}{2 \sigma}.
\end{equation}
The second one is the dynamical exponent $z$, which relates spatial and temporal correlation lengths:
\begin{equation}
z=\frac{\sigma}{\kappa}\,.
\end{equation}
We note that the same result for $z$ was obtained in the case of pure anomalous diffusion in space and time~\cite{Fogedby94}. The combined effect of spatial and temporal long-range interaction apparently predominates the critical dynamics to such an extent that the dynamical exponent $z$ locks onto the strict ratio $\sigma/\kappa$. 

To obtain the third exponent one has to consider the remaining flow equations
\begin{eqnarray}
\label{FlowEquations}
\partial_\ell \ln g    &=& 3\chi + d + z - L_g\\
\partial_\ell \ln a    &=& 2\chi + d + z - L_a
\end{eqnarray} 
in the vicinity of the fixed point
\begin{equation}
L_g^*=\frac{3\kappa\sigma-d\kappa-\sigma}{2\kappa}=\frac{\epsilon}{2},\qquad
L_a^*=\sigma\,,
\end{equation} 
where $\epsilon=d_c-d$ denotes the distance from the upper critical dimension
\begin{equation}
d_c = 3\sigma - \frac{\sigma}{\kappa}\,.
\end{equation}
To compute the matrix $M$ in Eq.~(\ref{MMatrix}) we note that

\begin{equation}
\frac{\partial (a L_a)}{\partial a} = \frac{L_g}{2}
\frac{\partial (a L_a)}{\partial g} = \frac{2 a L_a}{g}
\frac{\partial (g L_g)}{\partial g} = 3 L_g\,.
\end{equation} 
Here the first equation relies on the identity
\begin{eqnarray}
\frac{\partial}{\partial a} \Bigl[
G_0(\kvec',\omega')G_0(-\kvec',-\omega') \Bigr]
&=&
G_0(\kvec',\omega')G_0^2(-\kvec',-\omega')+\\
&&G_0^2(\kvec',\omega')G_0(-\kvec',-\omega') \nonumber
\end{eqnarray} 
relating the integrands of Eqs.~(\ref{LIntegral}) and~(\ref{VIntegral}). Again the off-diagonal entries of the matrix vanish to first order in $\epsilon$, hence to one-loop order the positive eigenvalue of $M$ is 
$\nu_\perp^{-1}=-(2\chi+d+z+\partial_a(aL_a))=\sigma-L_g^*/2=\sigma-\epsilon/4$. Summarizing these results, we therefore arrive at the critical exponents to one-loop order
\begin{equation}
\label{mainresult}
\beta = 1-\frac{\epsilon}{4 \sigma}+{\rm O}(\epsilon^2),\quad
\nu_\parallel = \frac{1}{\kappa}+\frac{\epsilon}{4 \kappa \sigma}+{\rm O}(\epsilon^2),\quad
\nu_\perp = \frac{1}{\sigma}+\frac{\epsilon}{4 \sigma^2} +{\rm O}(\epsilon^2).
\end{equation} 
%
%
\subsection{Phase diagram}
%
%
\begin{figure}
\centerline{\includegraphics[width=140mm]{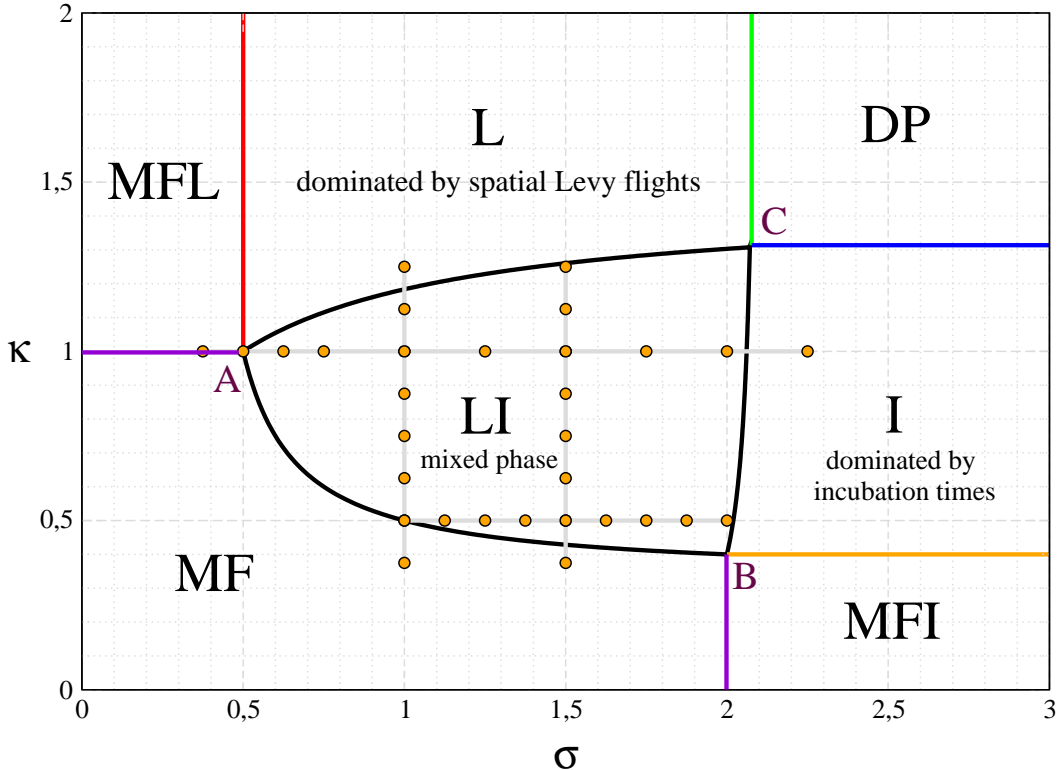}}
\caption{\label{FIGPHASEDIAG} \small
Phase diagram for epidemic spreading with spatio-temporal long-range interactions in one spatial dimension. The five phases and their boundaries are explained in the text. The circles along the grey lines indicate the points where the numerical simulations were carried out.}
\end{figure}

The field-theoretic results suggest a generic phase diagram in terms of the parameters $\sigma$ and $\kappa$. Fig.~\ref{FIGPHASEDIAG} shows the phase diagram in the case of one spatial dimension $d=1$, for which the numerical simulations are performed (see Sect.~\ref{numerical}). It comprises seven different phases, namely:
\begin{enumerate}
\item a mean field phase (MF) for small $\sigma$ and $\kappa$ which is governed both by L{\'e}vy flights \textit{and} incubation times. In this phase fluctuation effects are irrelevant and the exponents are given by $(\beta,\nu_\parallel,\nu_\perp)=(1,\kappa^{-1},\sigma^{-1})$, see Eq.~(\ref{MFExponents}).\\
\item a mean field phase for $\kappa>1$ governed solely by L{\'e}vy flights (MFL). In this regime waiting times described by the operator $\tilde{\partial}_t^\kappa$ are irrelevant compared to $\partial_t$ and the exponents are given by $(\beta,\nu_\parallel,\nu_\perp)=(1,1,\sigma^{-1})$.\\
\item a mean field phase for $\sigma>2$ dominated by incubation times (MFI). Here spatial L{\'e}vy flights associated with the operator $\tilde{\nabla}^\sigma$ become irrelevant and the exponents are given by $(\beta,\nu_\parallel,\nu_\perp)=(1,\kappa^{-1},1/2)$.\\
\item a directed percolation regime (DP) in the limit of short-range interactions (large $\sigma$ and $\kappa$) with the exponents
$\beta^{\rm DP} \simeq 0.2765$, $\nu_\perp^{\rm DP}\simeq 1.097$, and $\nu_\parallel^{\rm DP}\simeq 1.734$.\\
\item a non-trivial fluctuating phase dominated by spatial L{\'e}vy flights (L), where incubation times are still irrelevant so that the results of Refs.~\cite{JOWH99,HinrichsenHoward99} apply.  \\
\item a non-trivial fluctuating phase dominated by incubation times (I), where spatial  L{\'e}vy flights are irrelevant so that the results of Ref.~\cite{DalmaroniCardy} apply.\\
\item a non-trivial fluctuating mixed phase in the center (LI), on which we focus in the present work. In this phase spatial L{\'e}vy flights and incubation times are both relevant and the field-theoretic results derived in the previous subsection apply (see Eq.~(\ref{mainresult})).
\end{enumerate}

\noindent
Assuming that the critical exponents between the phases change continuously we can compute the phase boundaries by means of the scaling relations (\ref{TScaling}) and (\ref{SScaling}). For example, the spatial scaling relation (\ref{SScaling})  holds inside the phases L and LI while it is violated outside. Hence, plugging in the numerically known exponents of the adjacent phases we can compute phase boundaries in terms of $\sigma$ and $\kappa$. For example, inserting DP exponents into Eq.~(\ref{SScaling}) renders a vertical line at $\sigma=1+z^{\rm DP}-2\beta^{\rm DP}/\nu_\perp^{\rm DP}\simeq 2.077$
as a phase boundary. Similarly, inserting the mean field exponents (\ref{MFExponents}) into Eqs.~(\ref{TScaling}) and (\ref{SScaling}) gives the exact relation 
\begin{equation}
\kappa = \frac{\sigma} {3 \sigma -1}\,,
\end{equation}
which defines the curved phase boundary between the mixed and the mean field phase (line A-B in Fig.~\ref{FIGPHASEDIAG}). Finally, the remainig two curved phase boundaries of the mixed phase (lines A-C and B-C) are obtained by inserting the numerically estimated exponents for mere L{\'e}vy flights~\cite{HinrichsenHoward99} or mere incubation times~\cite{DalmaroniCardy} into the corresponding scaling relation.
   
In dimensions $d=2,3$  the area of the mixed phase (LI) is expected to shrink but the phase diagram still has the same qualitative structure. In $d > 4$ dimensions we expect the phases L,W, and LW to disappear so that the phase diagram shows only four different mean field phases separated by crossed straight lines at $\sigma=2$ and $\kappa=1$. 

%
%
\subsection{Critical behavior in the vicinity of $d_c=1$}
%
%
The predictive power of field-theoretic loop expansions is restricted to a small regime in the vicinity of the upper critical dimension. On the other hand numerical simulations of absorbing phase transitions are often carried out in one spatial dimension, i.e., usually far away from the regime where the field theory is valid. In the present model, however, the upper critical dimension depends continuously on the parameters $\sigma$ and $\kappa$ and thus we can tune them in such a way that $d_c$ is close to $1$. Therefore, we can quantitatively verify the field-theoretic predictions in 1+1-dimensional numerical simulation.

As an example we consider first-order corrections of the critical exponents when moving away from the mean field phase MF into the mixed phase LI along a horizontal line. For given $\kappa$ the line where $d_c=1$ is given by
\begin{equation}
\sigma_c = \frac{\kappa}{3\kappa-1}\,.
\end{equation} 
Let us now choose $\sigma=\sigma_c+\mu$ slightly above $\sigma_c$, where $0<\mu \ll 1$. To first order in $\mu$ the critical exponents are then given by
\begin{eqnarray}
\beta=1-\frac{(1-3\kappa )^2}{4 \kappa^2} \mu +{\rm O}(\mu^2)\,,\\
\nu_\parallel=\frac{1}{\kappa }+\frac{ (1-3 \kappa )^2}{4 \kappa^3}\mu+{\rm O}(\mu^2)\,,\\
\nu_\perp=3-\frac{1}{\kappa }-\frac{(\kappa +1)  (1-3\kappa )^2}{4 \kappa^3} \mu+{\rm O}(\mu^2)\,.
\end{eqnarray}
Similarly, for given $\sigma$ we have $d_c=1$ if $\kappa$ is equal to
\begin{equation}
\kappa_c = \frac{\sigma}{3\sigma-1}\,.
\end{equation} 
Choosing again $\kappa=\kappa_c+\eta$ slightly above $\kappa_c$, where $0<\eta \ll 1$ is the vertical distance from the phase boundary, the critical exponents are given to first order in $\eta$ by
\begin{eqnarray}
\label{BetaSeries}
\beta&=&1-\frac{(1-3 \sigma )^2}{4 \sigma^2}\eta +{\rm O}(\eta^2)\,, \\
\label{NuparSeries}
\nu_\parallel&=&3-\frac{1}{\sigma }-\frac{ (\sigma +1) (1-3\sigma )^2}{4 \sigma^3}\eta  +{\rm O}(\eta^2)\,,\\
\label{NuperpSeries}
\nu_\perp&=&\frac{1}{\sigma}+\frac{ (1-3 \sigma )^2}{4\sigma ^3}\eta  +{\rm O}(\eta^2)\,. 
\end{eqnarray}
%
%

\section{Numerical analysis}
\label{numerical}
%
\subsection{Seed simulations}
%
%
Modelling long-range interactions on a finite lattice almost inevitably evokes significant finite-size effects. Simulations studying the decay of a fully occupied state usually operate on a confined lattice, often with periodic boundaries. However, this approach requires huge lattice-sizes in order to minimize finite-size effects caused by long-range interactions, especially if $\sigma$ is small. As observed in Ref.~\cite{HinrichsenHoward99} this is already an issue when interactions are still local in time. Therefore we prefer so-called single-seed simulations~\cite{GrassbergerTorre}, where a cluster is generated by a single active site placed at the origin. Instead of a complete lattice, we now only store the individual coordinates of active sites. Thus, the lattice-size of the model is merely given by the machine limits for integer numbers ($\pm 2^{63}$), virtually eliminating finite-size effects.

In seed simulations one usually measures three time-dependent quantities, namely the survival probability $P_s(t)$, the current number of active sites $N(t)$, and their current mean square spreading from the origin $R^2(t)$. In most models with absorbing states, the survival probability scales as $P_s(t) \sim t^{-\delta}$, where $\delta=\beta/\nu_\parallel$. In the present model with incubation times, however, the survival probability is no longer well defined since there may be intermediate time intervals where the system is not active (see Ref.~\cite{DalmaroniCardy}). We therefore suggest to measure the autocorrelation function $P_a(t)$ which is defined as the probability of the site at $\xvec=0$ being active at time~$t$ and which remains well defined despite of incubation times. The autocorrelation function represents the full propagator of the field theory and is expected to scale as 
\begin{equation}
P_a(t)\sim t^{-2\delta}\,
\end{equation}
with $\delta=(d\kappa+\sigma-\kappa\sigma)/2\sigma$ inside the mixed phase.

The number of presently active sites $N(t)$ is measured as usual. This quantity is expected to scale as
\begin{equation}
N(t) \sim t^\theta\,,
\end{equation}
where $\theta=(d \nu_\perp-2\beta)/\nu_\parallel$. Inside regions I and LI, where incubation times are relevant, the additional scaling relation~(\ref{SScaling}) implies the simple exact expression
\begin{equation}
\theta=\kappa-1\,.
\end{equation}
Finally, the mean square spreading from the origin is expected to scale as
\begin{equation}
\label{RScaling}
{R^2}(t) \sim {t^{2/z}}\,,
\end{equation}
with $z=\sigma/\kappa$ inside the LI regime. However, the {\it arithmetic} average of $R^2(t)$ is problematic as it immediately diverges for $\sigma < 2$. As suggested in~\cite{HinrichsenHoward99} we therefore consider the {\it geometric} average instead, which is finite and turns out to comply with~(\ref{RScaling}).

Note that because of $\theta=-2\delta+d/z$ one of the three exponents is redundant. In fact, it would suffice to measure only two quantities (accounting for two scaling relations), but we deem it useful to have an additional numerical validation.
%
%
\subsection{Lattice model and its algorithmic implementation}
%
%
\begin{figure}
\centerline{\includegraphics[width=90mm]{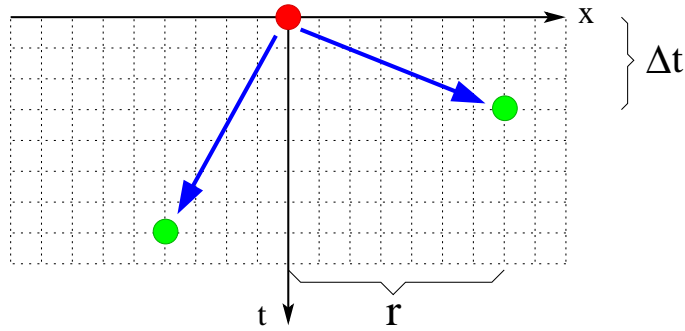}}
\caption{\label{FIGMODEL} \small
Lattice model used for the simulation. An active seed at the origin (red) infects two target sites (green) with probability $p$, spreading over a randomly generated spatial distance $r$ after an incubation time $\Delta t$ (see text). 
}
\end{figure}

As a generalization of directed bond percolation, our simulation features a 1+1-dimensional lattice 
with integer positions $x\in \mathbb{Z}$ (see Fig~\ref{FIGMODEL}). 
Starting with a single active seed located at $x = 0$, the infection
spreads as follows: For each active site at a given time, two random flight
distances are generated, one directed in positive, the other in
negative $x$-direction. The flight distances are distributed as
\begin{equation}
P(r) = \left\{ \begin{array}{ll} \sigma \:r^{-1-\sigma}
&\mbox{if}\   r > 1\\ 0 &\mbox{otherwise}\\ \end{array} \right.
\end{equation}
This distribution can easily be generated by drawing a random
number $q$ from a uniform distribution between 0 and 1 and setting
the flight distance $r=q^{-1/\sigma}$. Note that this implies a lower cutoff equal to the minimal flight distance of one lattice spacing.

Likewise, two random incubation times are generated, using a similar
probability distribution determined by $\kappa$:
\begin{equation}
P(\Delta t) = \left\{ \begin{array}{ll} \kappa \:\Delta t^{-1-\kappa}
&\mbox{if}\ \Delta t
> 1\\ 0 &\mbox{otherwise}\\ \end{array} \right.
\end{equation}
In both cases, fractional digits are truncated in order to
retrieve integer coordinates. Finally the two target sites are
independently infected with the probability $0 \leq p \leq 1$.

The spatial and temporal coordinates of active sites are stored in
a dynamically generated list. In fact, we use
a heap where we have immediate access to the next active site on
the time scale to be processed. Active sites with the
same time coordinate additionally use their spatial coordinate to maintain strict order. This allows the
active sites to be processed in proper sequence and hence tracking
down multiple activation easily (by definition each site can only be active once
at any given time).

\subsection{Simulation procedure}

Essentially, the model features three control parameters, namely
$\sigma$, $\kappa$ and $p$. To begin with, we pick out a point in
our phase diagram, specified by certain $\sigma$ and $\kappa$. Then we
determine the critical threshold ${p_c}$ for this pair of
control exponents. Averaging over numerous independent
realizations of the process, we measure the quantities $P_a(t)$, $N(t)$ and $R^2(t)$ as introduced above.
In order to determine the threshold probability ${p_c}$, we first
adjust $p$ such that we obtain asymptotically straight lines in a
double logarithmic plot. As usual, deviations from criticality
lead to a curvature of all the observed quantities in a double
logarithmic plot, $N(t)$ seeming to be the most sensitive to this.
We observe that the scaling relation $\theta = \kappa - 1$ holds
accurately inside the LI-phase (Fig.~\ref{FIGPHASEDIAG}). Therefore we then use
this scaling relation as a fit for $N(t)$ in order to improve the estimates of $p_c$.

%
\subsection{Numerical results}

%
%
\begin{figure}
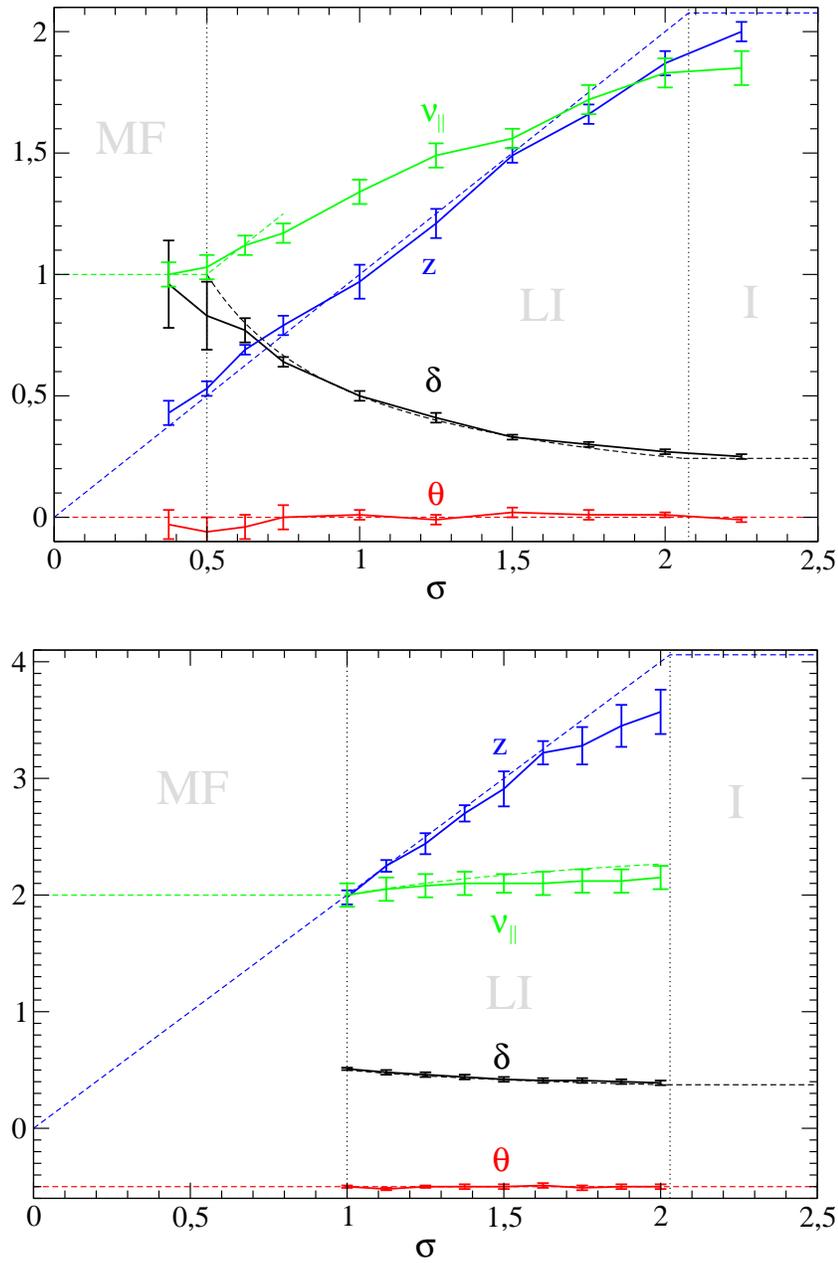

\centerline{\includegraphics[width=110mm]{cutt.eps}}
\vspace{6mm}
\centerline{\includegraphics[width=110mm]{cutb.eps}}
\caption{\label{CUTW} \small
Estimates of the critical exponents along the horizontal sections shown in Fig.~\ref{FIGPHASEDIAG}
for $\kappa=1$ (upper panel) and $\kappa=0.5$ (lower panel). Theoretical predictions are shown as dashed lines (see text).
}
\end{figure}

In order to obtain an overall impression, we opt for four different cross sections through the phase diagram, indicated as grey lines in Fig.~\ref{FIGPHASEDIAG}. Having determined the critical threshold for appropriate combinations of $\sigma$ and $\kappa$ along these cross sections, we measure $P_a(t)$, $N(t)$ and ${R^2}(t)$ at criticality. In a double logarithmic plot the asymptotic slopes of these quantities yield the exponents $-2\delta$, $\theta$ and $2/z$ respectively.

In order to verify the field-theoretic one-loop expansion, it is also necessary to study the off-critical regime. Choosing $p$ slightly below the critical threshold we again measure the average number of particles $N(t)$. According to standard scaling arguments this quantity should scale as
\begin{equation}
N(t) \sim t^\theta \, f\biggl(t \, (p_c-p)^{\nu_\parallel}\biggr)\,,
\end{equation}
where $f$ is a universal scaling function. Therefore, plotting $N(t)t^{-\theta}$ versus $t(p_c-p)^{\nu_\parallel}$ for different values of $p$ we can determine the exponent $\nu_\parallel$ by a data collapse. 

The numerical estimates of the critical exponents along the two horizontal sections through the mixed phase at $\kappa=1$ and $\kappa=1/2$ are shown in Fig.~\ref{CUTW}. The measured values are joined by solid lines which serve as a guide to the eye. The corresponding theoretical predictions are indicated by dashed lines. In the mixed phase LI the exponents  $\delta=(d\kappa+\sigma-\kappa\sigma)/2\sigma$, $\theta=\kappa-1$, and $z=\sigma/\kappa$ arise from the scaling relations and are expected to be exact. However, for the exponent $\nu_\parallel$ we can only apply the one-loop expansion derived in the previous section. For the horizontal section at $\kappa=1$ the predicted initial slope close to the mean field is again indicated by a dashed line. Since the cross section at $\kappa=1/2$ runs quite close to the mean field phase, we applied the one-loop expansion~(\ref{NuparSeries}) for the entire scope. Of course, with gradual departure from the phase boundary, the higher order corrections systematically increase the error of this approximation.

Yet, most of the numerical estimates are in fair agreement with the theoretical predictions. Systematic deviations beyond the error bars are observed in the crossover regime from the mixed to the incubation-dominated phase (LI$\to$I). Similar crossover effects were observed in previous numerical studies (see~\cite{HinrichsenHoward99}) and may be traced back to temporal finite-size effects. 

\begin{figure}
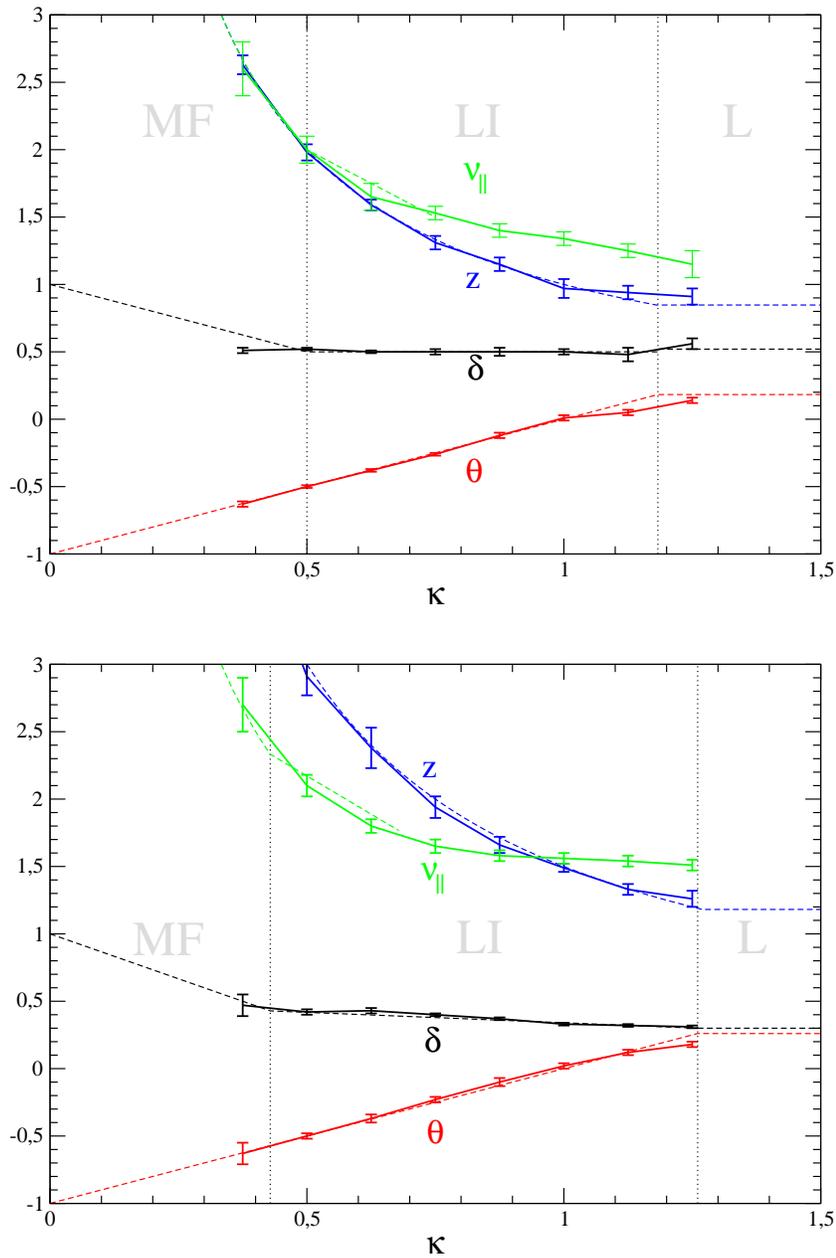

\centerline{\includegraphics[width=110mm]{cutl.eps}}
\vspace{6mm}
\centerline{\includegraphics[width=110mm]{cutr.eps}}
\caption{\label{CUTH} \small
Estimates of the critical exponents along the vertical sections at $\sigma=1$ (upper panel) and $\sigma=1.5$ (lower panel).
}
\end{figure}

Fig.~\ref{CUTH} shows the corresponding results for the two vertical sections at $\sigma=1$ and $\sigma=3/2$, respectively. In both cases the estimates for $\delta$, $\theta$ and $z$ are again in fair agreement with the theoretical predictions, showing systematic deviations in the crossover region between the phases LI and L. As for the exponent $\nu_\parallel$, the one-loop results~(\ref{NuparSeries}) again yield initial slopes close to the mean field which now coincide with the numerical estimates within error bars. 

To summarize, the combination of spatial and temporal long-range interactions in a spreading process is a numerically challenging task. Despite of the considerable numerical effort the precision of our estimates is limited. Nevertheless, we consider the numerical evidence as sufficient to validate the two scaling relations inside the mixed phase. Moreover, the estimates seem to be consistent with the one-loop calculation of the previous section. Because of crossover effects in finite-time simulations it is impossible to verify the exact location of the phase boundaries. Nevertheless, all results are consistent with the generic phase diagram presented in Fig.~\ref{FIGPHASEDIAG}.

\subsection{Subcritical behaviour}

As first noticed in~\cite{DalmaroniCardy} the effect caused by long-range interactions can also be perceived within the absorbing phase (i.e., where $p < p_c$ and thus the infection becomes extinct). Instead of an exponential decay, which would be typical for normal DP, the extinction of the epidemic is now asymptotically characterized by a power law. In fact, in the absorbing phase the density of infected sites decreases so rapidly that interactions by reinfection become irrelevant, i.e. the process is asymptotically dominated by sites that are infected once but then cease to newly infect other sites. Recalling that a single infection process complies with the distribution
\begin{equation}
\label{Distrib}
P(\Delta \xvec, \Delta t) \propto |\Delta \xvec|^{-d-\sigma} \Delta t^{-1-\kappa}\qquad |\Delta \xvec|, \Delta t \ge 1
\end{equation}
it is evident that the last active sites generated before the infection dies out give rise to a temporal distribution which is simply given by the temporal component of~(\ref{Distrib}). Integrating out the spatial component of~(\ref{Distrib}), one obtains power laws for $P_a(t)$ and $N(t)$ instead of exponential decays:
\begin{equation}
\label{Subcrit}
\begin{array}{l} P_a(t) \sim t^{-1-\kappa} \\ N(t) \sim t^{-1-\kappa} \\ \end{array} \qquad \mbox{if} \ p < p_c
\end{equation}
Note that the corresponding subcritical exponents become independent of the control parameters $\sigma$ and $p$.
 
Choosing $p$ close to $p_c$ one observes a prolonged crossover from an almost critical behaviour to the algebraic decay described by~(\ref{Subcrit}). As we discussed in the previous section, for $N(t)$ this crossover is characterized by a universal scaling function that involves $\nu_\parallel$. We validated these findings numerically at several points within the mixed phase (LI in Fig.~\ref{FIGPHASEDIAG}).

Interestingly, this type of asymptotic algebraic decay in the subcritical regime can be observed even within the DP phase (cf. Fig.~\ref{FIGPHASEDIAG}). This demonstrates again that the operators $\tilde{\nabla}^\sigma$ and $\tilde{\partial}_t^\kappa$ differ from their short-range counterparts $\nabla^2$ and $\partial_t$ for any finite $\sigma$ and $\kappa$ as they retain their long-range character. We note that this algebraic decay in the subcritical phase is not in contradiction with the DP hypothesis. 

\section{Conclusions}
\label{conclusions}
%
In this work we have studied the combined effect of spatial and temporal L{\'e}vy flights in a model of non-linear interaction. Field-theoretical considerations show us that only the free propagator is affected by the long-range terms, while the structure of the Feynman-diagrams remains unaltered. Furthermore the nonlocal operators which are responsible for L{\'e}vy flights and incubation times do not renormalize themselves, but instead renormalize their short-range counterparts. The calculation leads to two exact scaling relations which determine the three exponents $\delta=(d\kappa+\sigma-\kappa\sigma)/2\sigma$, $z = \sigma/\kappa$ and $\theta = \kappa - 1$. The only non-trivial exponent measured in the simulations is $\nu_\parallel$, which we calculated in the vicinity of the upper critical dimension to one-loop order.

For $\sigma = 2$, $\kappa = 1$ the long-range operators $\tilde{\partial}_t^\kappa$ and $\tilde{\nabla}^\sigma$ applied to a plane wave act in the same way as their short-range counterparts $\partial_t$ and $\nabla^2$. In a non-interacting model like anomalous diffusion~\cite{Fogedby94}, they would indeed be equivalent. However, it turns out that in an interacting theory, where fluctuations play an important role, long-range interactions are still relevant beyond this threshold. This issue is also evident from the location of the phase boundaries in the generic phase diagram.

To verify our theoretical results we have estimated the exponents along four cross sections through the phase diagram by numerical simulations in one spatial dimension.
Compared with the numerical results published almost ten years ago, the present estimates have about the same (sometimes even less) accuracy despite of a much higher numerical effort. However, in comparing these results one has to keep in mind that the combination of spatial and temporal L{\'e}vy flights is numerically much more challenging than the previously studied cases.
Nevertheless our numerical estimates are in fair agreement with the predictions of the field theory.

What can we learn from these results concerning epidemic spreading in nature? Recently Geisel and collaborators were able to show that individual dollar notes traced in space and time approximately move according to spatio-temporal L{\'e}vy flights with exponents $\sigma \approx \kappa \approx 1.5$~\cite{GeiselPrivate}, hence we have good reasons to expect that similar long-range effects will be relevant in the context of epidemic spreading. 

As a qualitative conclusion of the present study, we observe that long-range interactions generally decrease the critical threshold $p_c$ from where on the disease spreads. On the other hand, we find that the exponent $\beta$ increases with increasing interaction range, meaning that the divergency of the susceptibility with respect to the infection rate becomes less pronounced in the presence of L{\'e}vy flights, reaching a linear response ($\beta=1$) in the mean field limit. Thus we arrive at the conclusions that long-range transport mechanisms of modern societies such as aviation may facilitate the spreading of epidemics around the globe but at the same time the onset of spreading as a function of the infection rate becomes less dramatic, allowing for a better containment of the disease. However, realistic epidemic spreading is a highly inhomogenous process, smearing out the phase transition over a certain range, and therefore it is still an open question how robust these qualitative predictions are.

\vspace{5mm}
\noindent
\textbf{Acknowledgments:}\\
The numerical simulations were carried out on the 80-node Linux cluster an the University of W\"urzburg. We would like to thank A. Klein and A. Vetter for excellent technical support.

\appendix
\section{Tables of the numerical estimates}                               

\begin{tabular}{*{7}{>{$}r<{$}@{.}>{$}l<{$}}}

\toprule
\multicolumn{2}{c}{$\sigma$}
&\multicolumn{2}{c}{$\kappa$} &\multicolumn{2}{c}{$p_c$}
&\multicolumn{2}{c}{$\delta$} &\multicolumn{2}{c}{$\theta$}
&\multicolumn{2}{c}{$z$} &\multicolumn{2}{c}{$\nu_{\parallel}$}\\
\midrule

1&000&0&500&0&50146(2)&0&51(1)&-0&50(1)&1&98(6)&2&00(10)\\
1&125&0&500&0&50181(2)&0&48(2)&-0&51(1)&2&25(5)&2&05(10)\\
1&250&0&500&0&50218(2)&0&46(2)&-0&50(1)&2&44(9)&2&05(10)\\
1&375&0&500&0&50254(2)&0&44(2)&-0&50(2)&2&70(7)&2&05(10)\\
1&500&0&500&0&50289(1)&0&42(2)&-0&50(2)&2&91(14)&2&10(8)\\
1&625&0&500&0&50325(2)&0&40(2)&-0&49(2)&3&22(10)&2&10(10)\\
1&750&0&500&0&50354(2)&0&41(2)&-0&51(2)&3&28(16)&2&10(10)\\
1&875&0&500&0&50385(2)&0&40(2)&-0&50(2)&3&45(18)&2&12(10)\\
2&000&0&500&0&50416(2)&0&39(2)&-0&50(2)&3&57(19)&2&15(10)\\
\midrule

0&375&1&000&0&500595(3)&0&96(18)&-0&03(6)&0&43(5)&1&00(5)\\
0&500&1&000&0&501363(3)&0&83(14)&-0&06(6)&0&53(3)&1&03(5)\\
0&625&1&000&0&502598(3)&0&77(5)&-0&04(5)&0&69(2)&1&12(4)\\
0&750&1&000&0&504335(3)&0&64(2)&-0&00(5)&0&79(4)&1&17(4)\\
1&000&1&000&0&508815(3)&0&50(2)& 0&01(2)&0&97(7)&1&34(5)\\
1&250&1&000&0&513723(3)&0&41(2)&-0&01(2)&1&21(6)&1&49(5)\\
1&500&1&000&0&518405(3)&0&33(1)& 0&02(2)&1&49(3)&1&56(4)\\
1&750&1&000&0&522560(5)&0&30(1)&0&01(2)&1&66(4)&1&72(6)\\
2&000&1&000&0&526070(5)&0&27(1)&0&01(1)&1&87(5)&1&83(5)\\
2&250&1&000&0&528955(5)&0&25(1)&-0&01(2)&2&00(4)&1&85(7)\\
\midrule

1&000&0&375&0&50064(4)&0&51(2)&-0&63(2)&2&63(7)&2&6(2)\\
1&000&0&500&0&50146(2)&0&51(1)&-0&50(1)&1&98(6)&2&0(1)\\
1&000&0&625&0&502813(3)&0&50(1)&-0&38(1)&1&59(4)&1&65(10)\\
1&000&0&750&0&504630(5)&0&50(2)&-0&26(1)&1&31(5)&1&53(5)\\
1&000&0&875&0&506695(2)&0&50(3)&-0&12(2)&1&15(5)&1&40(5)\\
1&000&1&000&0&508815(3)&0&50(2)&0&01(2)&0&97(7)&1&34(5)\\
1&000&1&125&0&510855(5)&0&48(5)&0&05(2)&0&94(5)&1&25(5)\\
1&000&1&250&0&512744(8)&0&56(4)&0&14(2)&0&91(6)&1&15(10)\\
\midrule

1&500&0&375&0&50117(2)&0&47(8)&-0&63(8)&4&18(40)&2&7(2)\\
1&500&0&500&0&50289(1)&0&42(2)&-0&50(2)&2&91(14)&2&10(8)\\
1&500&0&625&0&505887(2)&0&43(2)&-0&37(3)&2&38(15)&1&80(5)\\
1&500&0&750&0&509780(5)&0&40(1)&-0&23(2)&1&94(8)&1&65(5)\\
1&500&0&875&0&514069(3)&0&37(1)&-0&10(3)&1&66(6)&1&58(4)\\
1&500&1&000&0&518405(3)&0&33(1)& 0&02(2)&1&49(3)&1&56(4)\\
1&500&1&125&0&522542(1)&0&32(1)&0&12(2)&1&33(4)&1&54(4)\\
1&500&1&250&0&5263297(3)&0&31(1)&0&18(2)&1&26(6)&1&51(4)\\

\bottomrule
\end{tabular}

\newpage

\vspace{5mm}
\noindent{\bf References}\\

\end{document}